\documentstyle[epsf,prl,aps]{revtex}

\newcommand{\tj}{$t$-$J$\ }

\begin{document}
\draft

 \twocolumn[\hsize\textwidth\columnwidth\hsize\csname @twocolumnfalse\endcsname

\title{Comment on ``Stripes and the \tj Model''}
\author{ Steven R.\ White$^1$ and D.J.\ Scalapino$^2$}
\address{ 
$^1$Department of Physics and Astronomy,
University of California,
Irvine, CA 92697
}
\address{ 
$^2$Department of Physics,
University of California,
Santa Barbara, CA 93106
}
\date{\today}
\maketitle


 ]

In a recent Letter\cite{hm} Hellberg and Manousakis (HM) studied a set of
16-site \tj clusters with two holes
using exact diagonalization to determine
whether the 2D \tj model has a striped ground state at a doping
of $x=1/8$. Based on these diagonalizations, they concluded that
the ground state of the 2D \tj model is uniform. They observed
low-lying nonuniform states with stripe-like features, 
but argue that these could only
represent ground state configurations if one applied
artificial boundary conditions. In particular, (1) they concluded
that there is ``no physical reason for such a simplified model
[the \tj model] to have a ground state with a periodic array of interfaces''; 
and (2) the striped states found in density matrix renormalization group (DMRG) 
calculations\cite{stripe,energetics,dmrg} are simply an ``artifact'' of the
boundary conditions used in DMRG.
We disagree with these conclusions and
believe that their analysis is flawed in several important
respects.

First, there is a physical mechanism which favors stripe formation in the
\tj model. As noted by HM and others \cite{PRD91},  for
$J/t$ in the relevant physical range, two holes on a \tj cluster will form
a pair with $d_{x^2-y^2}$ symmetry. Now the point is the pairs can lower
their energy further by forming a domain wall of holes with local pairing
correlations,  across which there is a
$\pi$-phase shift in the antiferromagnetic background \cite{KLB97}. 
Such an
arrangement reduces the frustration of the antiferromagnetic background
produced by the localized hopping of the
 holes and lowers their transverse kinetic energy, leading to
a stabilization of the domain walls. The details of the correlations
involved in this process have been studied extensively using DMRG and exact
diagonalizations \cite{PZ93,tprime,stripe,MEDpp}.
The second point raised by HM concerned the
question of boundary conditions. In using finite size clusters
to study models which may have broken-symmetry ground states, it
is often convenient to introduce a symmetry-breaking field and
then study the limiting behavior by first letting the size of
the system go to infinity and then letting the strength of the
perturbation go to zero. We view the open end boundary
conditions that we have used in this way and argue that far from
being artificial, they are important for understanding the
physics. Unfortunately, at present we are unable to carry out a
finite-size scaling analysis to obtain the infinite size
limit. Note that in the case of striped structures, the domain
wall spacing rather than the lattice spacing enters in setting
the lattice sizes required. Nevertheless, we have compared on
numerous occasions systems of different lengths, and not seen
any significant reduction in the stripe amplitudes. 
We have also compared a
$12\times6$ system with the interior $12\times6$ region of
a $24\times6$ system at the same doping where we found the energy
per hole to be the same to within about $\pm 0.01t$. 
Note also that
we see the stripes regardless of whether we apply a staggered
field to the edges of the system. In the absence of the staggered field,
it simply takes 
longer for the calculation to converge.
Finally, we {\it are}  able to observe an essentially uniform ground
state even with open boundary conditions; they occur when
a next-nearest neighbor hopping $t'$ is made large
enough ($t' \sim 0.3 t$) \cite{tprime}. The effect of this term
is to destabilize the domain walls and favor a gas of pairs.

We also see stripes develop (without $t'$)
as our DMRG calculation progresses from a starting point with
all the holes in a clump in the center of a long system; the
stripes appear spontaneously 
long before the holes have any probability amplitude 
of being near the open ends of the system. Such a calculation
is shown in Figs. 1 and 2. Here, a $16\times6$ system with $J/t=0.35$
and cylindrical boundary conditions, with eight holes,
is studied with DMRG. No external fields were applied.
In the initial DMRG sweep, all the holes
were forced onto the center two columns of sites. Subsequently,
as the finite system DMRG sweeps are performed, the system moves
the holes in order to decrease the energy of the wavefunction.
Since hole density is locally conserved, the essentially local
DMRG sweeps move the holes slowly.
N\'eel order develops spontaneously in the $z$ spin direction,
since we have quantized the spins in the $z$ basis. As the
calculation converges, this spontaneously broken spin symmetry
slowly disappears, corresponding to an averaging of the overall spin
direction over all possible directions. (This reduction in the
local spin moments is not yet visible in the sweeps shown in
Fig. 1.)
Substantially before the hole density has approached either end
of the system, two stripes appear spontaneously. The $\pi$ phase
shift also appears spontaneously, and is visible in the local
measurements because of the broken spin symmetry. As the
calculation converges to the ground state, the $\pi$ phase shift
becomes visible only through spin-spin correlations.
The two stripes repel, and continue to move slowly apart as
the sweeps progress until they are roughly equidistant from each
other and the open left and right ends.

Turning now to the calculations reported by HM,
first, it appears that by searching for the lowest energy
states in a variety of 16 site clusters, they have simply found
the clusters with the largest finite size effects. In this
regard, it is
interesting to note that the lowest-energy labeled states (a), (b), and
(c) are all on the most one-dimensional clusters of the ones they studied,
 those with
one of the primitive translation vectors being (2,2). The
fourth lowest energy state (d) is on the next most one-dimensional
lattice. The authors excluded clusters which were even more one dimensional
than these because of finite size effects;
if they had also excluded the (2,2) clusters, their conclusions
would have been quite different.  The important point is that
 the energy differences they
obtain by comparing different small clusters
are far {\it too large} to relate to the {\it subtle competition} between
pairing and stripes.  Their lowest energy
of $-0.660t$ per site translates to an energy of $-1.979t$ per
hole. In an exact diagonalization, we find that the ground state
energy of the ordinary $4\times4$ periodic cluster with two holes is $-0.628t$
per site, or $-1.72t$ per hole. From this we see that none of
the states shown in Fig. 1 (which unfortunately are not labeled
according to cluster) is from the $4\times4$ cluster---they were
omitted by HM because their energy was too high.
This difference of $0.26t$ per hole between these two uniform
ground states of different 16 site clusters is about an order of
magnitude larger than the energy
difference between pairs and stripes on large systems at low doping
shown in Fig.~1 of Ref.~\cite{stripe}.

Secondly, the stripes we have found involve correlations between two or
more pairs of holes, while the 16-site clusters studied by Hellberg and
Manousakis have only one pair. As noted earlier by Prelovsek and Zotos
\cite{PZ93} one
needs at least 4 holes (i.e.~two pairs) on a cluster to study stripe
formation. In fact, using diagonalizations of clusters containing 4 holes,
Prelovsek and Zotos\cite{PZ93} 
found evidence for domain wall formation. We believe
that meaningful information about stripe stability cannot come from
calculations which involve only one pair on a cluster. In fact, 
the ``striped'' states reported in the HM Letter have
quite a different origin than the stripes in the many-hole systems we have
found. Just as one can combine $p$ and $-p$ excited states of a single
particle in a periodic box to create a spatial density wave 
oscillation, these authors have combined degenerate excited states of one
pair in a cluster to create a standing density wave. The stripes we have
found arise from correlations of pairs rather than the excited state of one
pair.

To summarize: there is a physical mechanism for domain wall
formation in the \tj model and the boundary conditions used in
our calculations are not artificial, but, in fact, provide a
simple way of introducing a symmetry breaking field. We have
seen the stripes disappear when we change the model by adding a
nearest-neighbor hopping $t'$, showing that the model itself does
contain a mechanism for stripe formation. We have seen little
change in behavior in going to larger lattices, although we have
not been able to carry out a finite size scaling analysis.
We believe that the calculations reported by HM 
are misleading because of large finite size effects
on their 16-site clusters and the fact that they only have one
pair. In particular, the stripe-like patterns they observe are only 
standing wave patterns of the motion of a pair of holes rather than the
stripes arising from correlations of many pairs of holes (e.g.~in the 
$16\times 8$ cluster of reference \cite{stripe} there were 16 holes or 8
pairs).
Finally, it is important to note that we
have not argued that stripes in the 2D \tj model are necessarily
static, although they certainly are in our calculations. 
There may be low energy fluctuations of the stripes
which restore translational or rotational symmetry, which 
require larger systems and higher accuracy than we currently can
manage. However, what we do believe our calculations show is
that a ``uniform'' many-hole state which has no manifestation of static
or dynamic stripes is, in fact, {\it not}  a low-lying state of
the \tj model near a doping $x=0.125$.
Indeed, most of the
experiments which see stripes find dynamic
stripes \cite{stripeexp}; they
are observable in dynamical susceptibilities, regardless of
whether there are broken spatial symmetries.


S.R.~White acknowledges support from the NSF under
grant \# DMR98-70930 and D.J.~Scalapino acknowledges support
from the NSF
under grant \# DMR95-27304.

\newpage
\begin{figure}[ht]
\epsfxsize=3.0 in\centerline{\epsffile{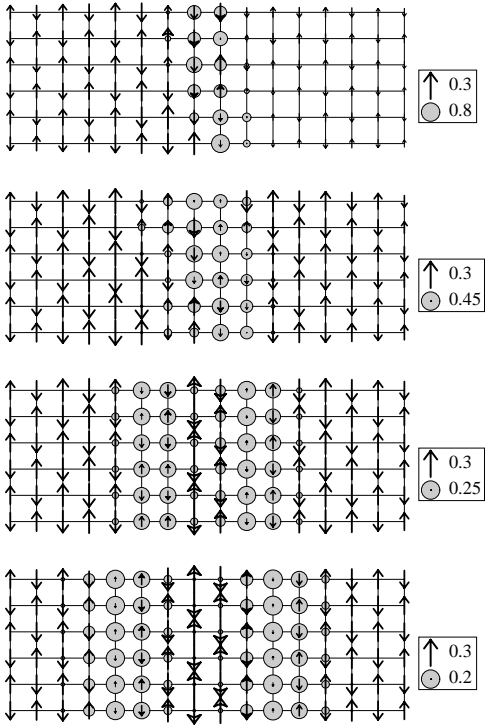}}
\caption{
A $16\times6$ \tj system, with $J/t=0.35$ and eight holes,
with cylindrical boundary conditions (open in the $x$ direction,
periodic in $y$), 
is studied with DMRG. The plot shows the local hole density by
the diameter of the circles, according to the scales shown.
The local spin moment is shown by the length of the arrows.
The four pictures represent the state of the system at the end
of sweeps 1, 3, 6, and 15. The number of states kept per block
was increased as the sweeps progressed, with 80, 200, 600, and
1000 states kept in these four sweeps, respectively.
The energy steadily decreased, taking values -42.96, -49.68,
-51.890, and -52.279, in the four sweeps.
The quantum numbers during the DMRG
warmup sweep (sweep ``0'') were manipulated to force all 8 holes onto the
center two columns, in order to strongly favor a phase-separated state.
However, the phase separated state is unstable, and splits into
two four-hole stripes, which subsequently repel each other.
Different style arrows are used to distinguish the two separate
antiferromagnetic domains that form.
}
\label{figone}
\end{figure}

\newpage
\begin{figure}[ht]
\epsfxsize=3.0 in\centerline{\epsffile{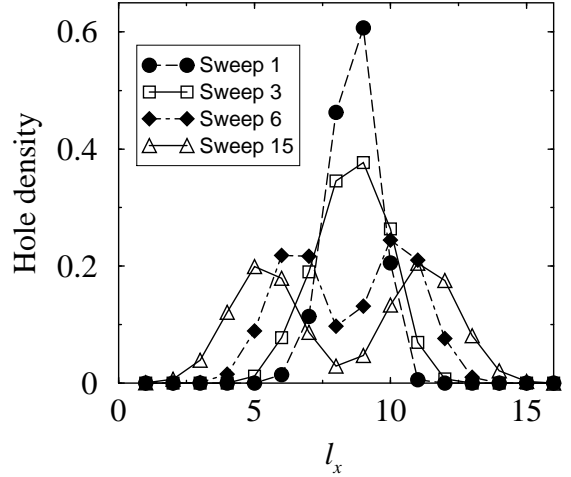}}
\caption{
The hole density as a function of the $x$ coordinate is
shown for the same four sweeps as in Fig. 1. Note that in sweep 6,
where the striped pattern is clearly visible, the hole density on the left
and right edges is still zero. Consequently, the stripes 
are not caused by the open boundaries. However, we find that
whether the final stripe configurations are site centered or 
bond centered is influenced by the boundary conditions.
}
\label{figtwo}
\end{figure}

\end{document}